\documentclass[a4paper,fleqn]{cas-dc}

\usepackage[numbers,sort&compress]{natbib}
\usepackage{siunitx}

\def\tsc#1{\csdef{#1}{\textsc{\lowercase{#1}}\xspace}}
\tsc{WGM}
\tsc{QE}
\tsc{EP}
\tsc{PMS}
\tsc{BEC}
\tsc{DE}
\begin{document}
\let\WriteBookmarks\relax
\def\floatpagepagefraction{1}
\def\textpagefraction{.001}
\shorttitle{Insights into ventilation hysteresis shift due to flow unsteadiness in ventilated supercavitation}
\shortauthors{Yoon et~al.}

%\title [mode = title]{Effect of strut and cavitator shape on the ventilation demand of supercavitation in unsteady flows}
\title [mode = title]{Insights into ventilation hysteresis shift due to flow unsteadiness in ventilated supercavitation}
%\tnotemark[1,2]

%\tnotetext[1]{This document is the results of the research
%   project funded by the National Science Foundation.}

%\tnotetext[2]{The second title footnote which is a longer text matter
%   to fill through the whole text width and overflow into
%   another line in the footnotes area of the first page.}

\author[1,2]{Kyungduck Yoon}[%type=editor,
                        %auid=000,bioid=1,
                        %prefix=Sir,
                        %role=Researcher,
                        %orcid=0000-0001-7511-2910
                        ]
%\cormark[1]
%\fnmark[1,2]
%\ead{yoon0108@umn.edu}
%\ead[url]{www.cvr.cc, cvr@sayahna.org}

%\credit{Conceptualization of this study, Methodology, Software}

\address[1]{St. Anthony Falls Laboratory, University of Minnesota, Minneapolis, MN 55414, USA}
\address[2]{The George W. Woodruff School of Mechanical Engineering, Georgia Institute of Technology, Atlanta, GA 30332, USA}
\address[3]{Department of Mechanical Engineering, University of Minnesota, Minneapolis, MN 55414, USA}
\address[4]{Department of Mechanical Engineering, School of Engineering, University of Petroleum and Energy Studies, Energy Acres, Bidholi, Dehradun, Uttarakhand 248007, India}

\author[1,3]{Jiaqi Li}[
%style=chinese
]

\author[1,3]{Siyao Shao}[%
   %role=Co-ordinator,
   %suffix=Jr,
   ]
\author[4]{Ashish Karn}[%
   %role=Co-ordinator,
   %suffix=Jr,
   ]
\author[1,3]{Jiarong Hong}[
]
\cormark[1]

\ead{Jhong@umn.edu}

\cortext[cor1]{Corresponding author}
%\cortext[cor2]{Principal corresponding author}
%\fntext[fn1]{This is the first author footnote. but is common to third
%  author as well.}
%\fntext[fn2]{Another author footnote, this is a very long footnote and
%  it should be a really long footnote. But this footnote is not yet
%  sufficiently long enough to make two lines of footnote text.}

%\nonumnote{This note has no numbers. In this work we demonstrate $a_b$
%  the formation Y\_1 of a new type of polariton on the interface
%  between a cuprous oxide slab and a polystyrene micro-sphere placed
%  on the slab.
%  }

\begin{abstract}
Understanding the air injection strategy of a ventilated supercavity is important for designing high-speed underwater vehicles wherein an artificial gas pocket is created behind a flow separation device to reduce skin friction. Our study systematically investigates the effect of flow unsteadiness on the ventilation requirements to form ($C_{Qf}$) and collapse ($C_{Qc}$) a supercavity. Imposing flow unsteadiness on the incoming flow has shown an increment in higher $C_{Qf}$ at low free stream velocity and lower $C_{Qf}$ at high free stream velocity. High-speed imaging reveals distinctly different behaviors in the recirculation region for low and high freestream velocity under unsteady flows. At low free stream velocities, the recirculation region formed downstream of a cavitator shifted vertically with flow unsteadiness, resulting in lower bubble collision and coalescence probability, which is critical for the supercavity formation process. The recirculation region negligibly changed with flow unsteadiness at high free stream velocity and less ventilation is required to form a supercavity compared to that of the steady incoming flow. Such a difference is attributed to the increased transverse Reynolds stress that aids bubble collision in a confined space of the recirculation region. $C_{Qc}$ is found to heavily rely on the vertical component of the flow unsteadiness and the free stream velocity. Interfacial instability located upper rear of the supercavity develops noticeably with flow unsteadiness and additional bubbles formed by the distorted interface shed from the supercavity, resulting in an increased $C_{Qc}$. Further analysis on the quantification of such additional bubble leakage rate indicates that the development and amplitude of the interfacial instability accounts for the variation of $C_{Qc}$ under a wide range of flow unsteadiness. Our study provides some insights on the design of a ventilation strategy for supercavitating vehicles in practice.

\end{abstract}

%\includegraphics{figs/grabs.pdf}

\begin{highlights}
\item Formation ventilation correlated with behavior of cavitator recirculation zone
\item Recirculation position and size vary upon flow unsteadiness
\item Formation ventilation can decrease upon increasing Fr and flow unsteadiness
\item Instability of cavity interface correlated with collapse ventilation demand
\end{highlights}


\maketitle

\section{Introduction}
\label{intro}
Ventilated supercavitation refers to the formation of an artificial gas pocket in water flow created by air injection behind a flow separation device, i.e., cavitator in such a way that the so-formed cavity is large enough to surround an immersed vehicle. 
This phenomenon has been broadly investigated for its potential applications in the drag reduction for high-speed operation of underwater vehicles \cite{franc2006book}. 
Due to the complex multi-phase interactions involved in cavitating flows, which are sensitive to flow conditions, a significant amount of research has been conducted on characterizing the behaviors of ventilated supercavities \cite{may1975report} as well as on mechanical control strategies \cite{escobar2015thesis}. 
However, despite numerous research reported on the characterization of general behaviors of ventilated supercavities \cite{Nesteruk2012Book}, an area that has not hitherto received significant attention, is the ventilation strategy, i.e. optimally controlling the ventilation rate for the supercavity to be formed and sustained under various flow conditions. 

Compared to several investigations that report general cavity behaviors, such as geometry, shape, and cavity closure, only a handful of studies focus on the ventilation requirements associated with the supercavity formation and its sustenance upon formation. 
For instance, Karn et al. \cite{karn2016aJFM}, explored the ventilation hysteresis phenomenon in great detail and established that the ventilation demands to form and to sustain a supercavity may be significantly different, the latter being much smaller than the former. 
In a followup work, Karn et al. \cite{karn2016bETFS} investigated the ventilation demands of the supercavity under various flow settings and provided a detailed explanation of the cavity formation and collapse processes, relating each with bubble coalescence proficiency and pressure balance near the closure. 
Their research has been conducted for a backward-facing model (BFM) with only a disk type cavitator. 
However, there is an inherent limitation with this type of cavitator configuration – it neglects the effects of the cavitator shape or the presence of the mounting strut, both of which have been shown to noticeably affect the cavity behaviors \cite{semenenko2001report, ahn2010IJNAOE, moghimi2017JAFM}, especially the supercavity formation process \cite{ahn2018CAV}. 
Recently, Shao et al. \cite{shao2020ETFS} studied the ventilation demands for a forward-facing model (FFM) with a variety of cavitator geometries (cone, disk, and non-axisymmetric) to consider both the cavitator shape and the mounting strut effects. They have found out that the cone type cavitator required the least ventilation flow to form and sustain a cavity. 
They have also observed that the interaction between the mounting strut and the air-water interface leads to a noticeable change in collapse ventilation demand compared to that of the BFM case. 
Their detailed analysis on the momentum balance between the air injection and the estimated re-entrant jet at the closure further supported that the re-entrant jet governs the cavity collapse process. 
However, although it has been reported that the ventilation demand depends crucially on the flow unsteadiness \cite{karn2016bETFS}, which may significantly alter the operation of the supercavitating object \cite{lee2013JFE,escobar2015thesis}, the investigations on ventilation demand and ventilation hysteresis to date have been limited to the steady flow conditions only. 
Therefore, to connect the lab-scale experiments with the practical situations of underwater vehicles encountering surface waves, experimental investigations exploring the role of different cavitator shapes and mounting strut effects in unsteady flows is needed, not only to understand general cavity behaviors, but to investigate the underlying physics with an express emphasis on the ventilation demand and ventilation hysteresis.

Though mostly limited to the general cavity behaviors or the ventilation demands of the BFM supercavity, a few recent studies investigated ventilated supercavitation under unsteady flows by using a gust generator that consists of flapping hydrofoils \cite{lee2013JFE,karn2015ETFS,karn2016bETFS,shao2018POF}. 
Such a setup was deployed to simulate unsteady incoming flows by controlling either the angle of attack ($AoA$) or the flapping frequency ($f_g$) of the hydrofoils. 
In particular, it has been reported that for the unsteady flows the cavity dimensions and cavitation number ($\sigma_c$) periodically change \cite{lee2013JFE} and closure variation is observed between twin-vortex and re-entrant jet \cite{karn2015ETFS}. 
Shao et al. \cite{shao2018POF} further classified FFM supercavity into five distinct states (namely stable, wavy, pulsating 1, pulsating 2, and collapsing states), characterizing each state based on the simultaneous pressure measurement and high-speed imaging. 
They observed transitions across these states with a change in either $AoA$ or $f_g$ and further proposed a stability criterion for these state transitions. 
Karn et al. \cite{karn2016bETFS} studied the formation and collapse ventilation demand trends with respect to the change in $AoA$ and $f_g$ for BFM and observed that all these demands increase with higher flow unsteadiness. 
They noted that such flows impose a vertical perturbation to the individual bubble movements that lead to the increased formation ventilation demand. 
They also commented on the internal pressure fluctuation that leads to higher collapse ventilation demand. 
However, their explanations of the observed trends heavily relied on the implications of flow perturbation and pressure fluctuation but lacked visual evidence. 

Therefore, as a follow-up study of Shao et al. \cite{shao2020ETFS}, we present a systematic examination of unsteady flow conditions, generated by the gust generator, on the formation and sustenance ventilation demand of FFM model (including the mounting strut effect) with a cone type cavitator which has shown to require the least ventilation demand to form and sustain the supercavity. 
The rest of the sections are as follows: Section~\ref{sec:Methods} provides detailed explanation on the experimental methods. The results of our study are presented in Section~\ref{sec:Results}. Specifically, Section~\ref{sec:lowfr}-~\ref{sec:highfr} demonstrates the results of cone cavitator with different flow regimes. Section~\ref{sec:Shed} provides a quantitative estimation of change in the collapse ventilation demand due to the flow unsteadiness.  Section~\ref{sec:Conclusions} provides a summary of the current study.

\begin{table}[width=1\linewidth,cols=3,pos=b]
\caption{Experimental conditions illustrating the incoming flow unsteadiness. Froude number ($Fr$) is calculated based on the cavitator diameter ($d_c$). In the table, $Fr\ 5$ refers to low $Fr$ and $Fr\ 15$ refers to high $Fr$ hereafter, which is demarcated based on the bubble concentration regime from previous studies \cite{karn2016bETFS,shao2020ETFS}.}\label{table1}
\begin{tabular*}{\tblwidth}{@{}CCC@{} }
\toprule
$Fr=U/\sqrt{gd_c}$ & $AoA\ [\si{\degree}]$ & $f_g\ [\mathrm{Hz}]$\\
\midrule
$ $ & $0$ & $0$ \\
$5$ & $2,6,10$ & $1$ \\
$ $ & $2,6,10$ & $5$ \\
 \midrule
$ $ & $0$ & $0$ \\
$15$ & $2,6,10$ & $1$ \\
$ $ & $2,6,10$ & $5$ \\
\bottomrule
\end{tabular*}
\end{table}

\section{Experimental Setup and Methodology}\label{sec:Methods}

\begin{figure}
  \centering
  \includegraphics[scale=1]{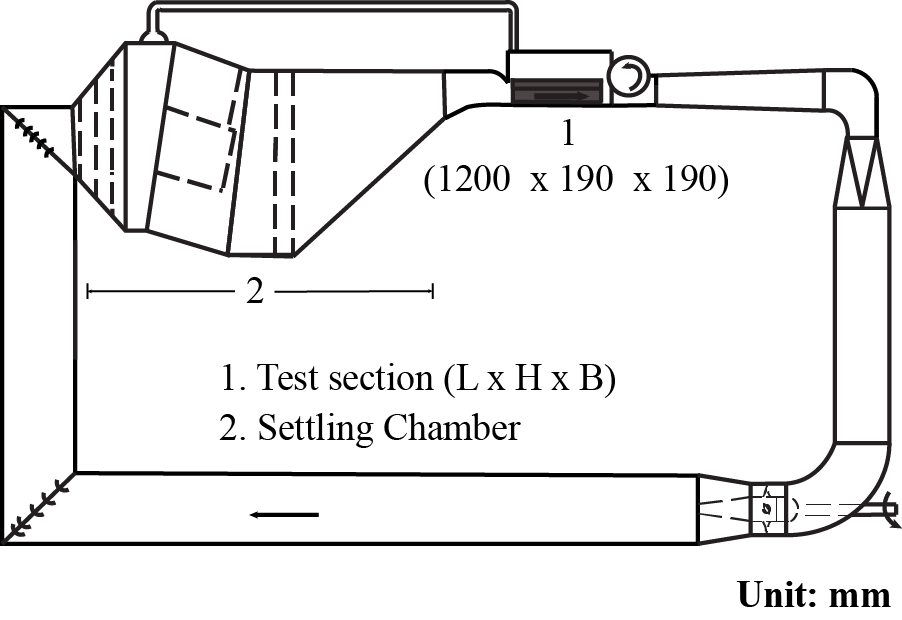}
  \caption{High-speed cavitation water tunnel at the Saint Anthony Falls Laboratory (SAFL), University of Minnesota. This schematic is adapted from \cite{shao2020ETFS}.}
\label{fig1}
\end{figure}

The experiments are conducted in the high-speed cavitation water tunnel at Saint Anthony Falls Laboratory (SAFL), University of Minnesota. 
As shown in Fig.~\ref{fig1}, the flow facility consists of a closed recirculating tunnel with a large volume dome-shaped settling chamber located upstream of the test section, designed for fast bubble removal during the ventilation experiments. 
The dimension of the test section is $1.20\ \mathrm{m}\ \times\ 0.19\ \mathrm{m}\ \times\ 0.19\ \mathrm{m}$ (length, height, and width), and the bottom and the two side windows of the test section are made up of Plexiglas for optical access. 
During the experiments, the free stream velocity is calculated through a Rosemount 3051 differential pressure sensor that measures the differential pressure between the settling chamber and the test section. 
Correspondingly, the desired free stream velocity in the test-section is regulated by feedback control of the motor attached to the centrifugal pump, located at the bottom section of the water tunnel facility. 
In recent years, this facility has been used extensively for a number of ventilated partial- and supercavitation experiments \cite{jiang2018POF,wu2019JFM,yoon2020EXIF}.

\begin{figure}
  \centering
  \includegraphics[scale=1]{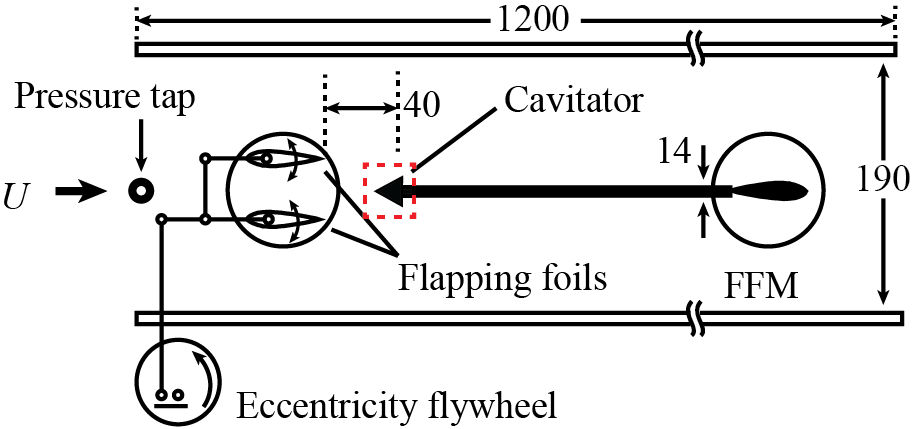}
  \caption{Test section with gust generator setup. Units are in $\mathrm{mm}$.}
\label{fig2}
\end{figure}

Similar to the setup from \cite{shao2018POF}, the gust generator is installed at the upstream of the test section (Fig.~\ref{fig2}). 
Unsteady flows are generated by the continuously flapping NACA0020 hydrofoils (detailed in \cite{kopriva2008JFE}) with various range of $AoA$ and $f_g$ imposing a vertical perturbation in the flow that propagates downstream. 
During the experiments, supercavitation is generated by ventilating air behind the cavitator. The mass flow rate of the ventilated air is controlled by an Omega Engineering FMA-2609A mass flow controller that has a proportional-integral-derivative algorithm for controlling at a constant desired flow rate up to 55 SLPM with the uncertainty within $\pm0.2\%\ FS$. 
A FFM cone-type cavitator with a diameter of 30 mm has been employed, which is the exact same model as reported by \cite{shao2020ETFS} to facilitate the investigation of unsteady flow effects.

Table~\ref{table1} lists the conditions that are investigated in the current experiment. 
Froude number ($Fr$) is used to scale the free stream velocity, with a characteristic length of the cavitator diameter ($d_c$). 
The experiments are conducted at two different $Fr$, and in each case the $AoA$ and $f_g$ are varied in a wide range to simulate various incoming flow unsteadiness. 
The underlying reasons for the choice of such Fr will be discussed shortly hereafter. 
The steady flow condition for comparison refers to zero degree of $AoA$ and zero frequency of $f_g$, i.e., foils parallel to the unperturbed flow direction. 
During the experiments, the ventilation demands ($C_Q=\dot{Q}/Ud_c^2$) are measured five times at each flow condition to ensure robustness from potential outliers. 
Specifically, the ‘Formation ventilation demand’ ($C_{Qf}$) refers to a critical ventilation rate at which the bubbles beyond the cavitator start to coalesce and form a stable supercavity with a transparent air-water interface. 
The ‘Collapse ventilation demand’ ($C_{Qc}$) refers to the minimum ventilation rate required, at which a supercavity can be sustained upon formation, without collapsing. 
The measurements of these ventilation demands are recorded by changing the ventilation rates with an increment or decrement less than $0.05$ SLPM to ensure a stable variation of ventilation rate and to accurately capture the critical value of $C_{Qf}$ and $C_{Qc}$.

\begin{figure}%[pos=b]
  \centering
  \includegraphics[scale=1]{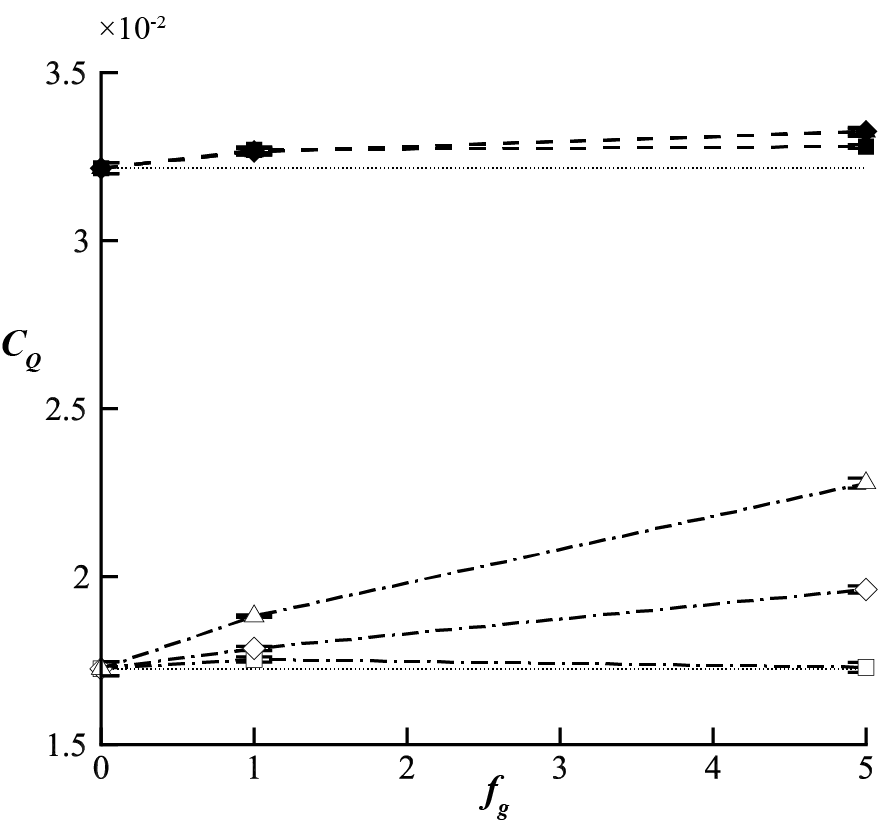}
  \caption{Ventilation coefficients for low $Fr$ regime under varying flow unsteadiness modified by $AoA$ and $f_g$. Dotted horizontal lines indicate steady flow condition, dashed lines indicate $C_{Qf}$ trends, dashed-dotted lines indicate $C_{Qc}$ trends. Symbols: $\blacksquare$, $C_{Qf}$ at $AoA\ 2\si{\degree}$; $\square$, $C_{Qc}$ at $AoA\ 2\si{\degree}$; $\blacklozenge$, $C_{Qf}$ at $AoA\ 6\si{\degree}$; $\lozenge$, $C_{Qc}$ at $AoA\ 6\si{\degree}$; $\blacktriangle$, $C_{Qf}$ at $AoA\ 10\si{\degree}$; $\triangle$, $C_{Qc}$ at $AoA\ 10\si{\degree}$.}
\label{fig3}
\end{figure}

The ventilation demands are investigated for $Fr$ $5$ (low) and $15$ (high) that represents different regimes of bubble size and concentration \cite{karn2016bETFS,shao2020ETFS}. 
Particularly, the low $Fr$ regime, also referred to as the low bubble concentration regime, consists of a relatively bigger bubble size distribution \cite{karn2016cETFS}, and increasing flow speed in this regime breaks up individual bubbles. 
Therefore, in the low Fr regime, increasing $Fr$ results in the growth of $C_{Qf}$ since more ventilation is required to increase the size and number of bubbles to coalesce and form a supercavity \cite{karn2016bETFS}. 
High $Fr$ regime, on the contrary, consists of a higher concentration of smaller bubble size due to the dominant role of turbulence-induced bubble breakup. 
Consequently, in high $Fr$ regime, an increase in flow speed favors the bubble coalescence process and therefore $C_{Qc}$ decreases, since the bubble concentration in this regime is already high enough to restrict the free space for bubble movement and eventually increase the chance of bubble collision \cite{karn2016bETFS}. 
A critical $Fr$ that demarcates such different regimes is found to be around 10 \cite{karn2016bETFS,shao2020ETFS}.

\begin{figure*}
  \centering
  \includegraphics[scale=1]{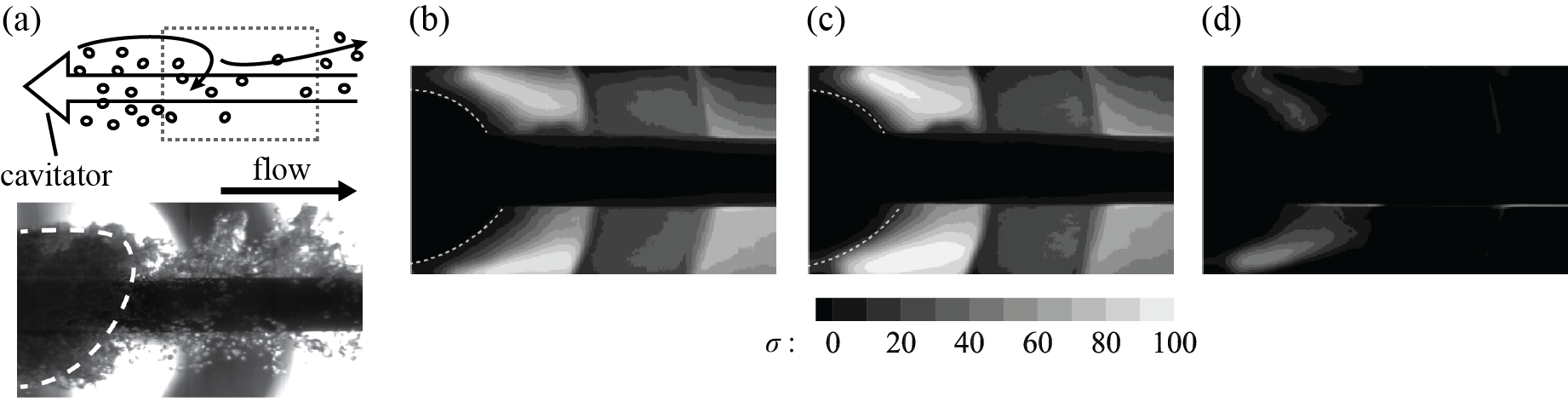}
  \caption{(a) Sample high-speed image at the field-of-view (highlighted region in the schematics) is taken at $C_Q$ slightly below $C_{Qf}$. The region marked with the white dotted line in the sample image indicates the recirculation region with constantly low intensity due to bubbles blocking the light path from the illumination source. Based on the image sequence, standard deviations of image intensity for the (b) steady and (c) unsteady flows ($AoA\ 6\si{\degree},\ f_g\ 5\ \mathrm{Hz}$) at low $Fr$ are plotted at each pixel location. The brighter region indicates a higher fluctuation of the brightness throughout the recorded sequence.  Regions marked with gray dotted lines imply the size of the recirculation region for the steady flow case. The difference between steady and unsteady flows are plotted in (d), indicating that the incoming flow unsteadiness imposes vertical fluctuation of the recirculation region.}
\label{fig4}
\end{figure*}

A Photron APX-RS high-speed camera is deployed to furnish visual evidences on the observed trends of $C_Q$. 
The time resolution for the high-speed imaging varies with $Fr$, from $3000\ \mathrm{Hz}$ (for low $Fr$ regime) to $9000\ \mathrm{Hz}$ (for high $Fr$ regime) with a sensor size of $512\times 512$ pixels. 
The high-speed images for $C_{Qf}$ are taken at the foamy cavity state (terminology from \cite{karn2016aJFM}) with the ventilation rate slightly below $C_{Qf}$ to infer the effect of flow unsteadiness in the bubble coalescence process. 
For investigating $C_{Qc}$, the high-speed images are acquired at the ventilation rate slightly above $C_{Qc}$ to infer the effect of flow unsteadiness on the gas leakage mechanism and the cavity stability.

\section{Results}\label{sec:Results}

\subsection{Ventilation demands in low $Fr$ regime}\label{sec:lowfr}

As shown in Fig. 3, $C_Q$ at low $Fr$ regime are plotted from steady to various unsteady incoming flow conditions. 
In this regime, $C_{Qf}$ trends (dashed lines) show a slight increase with $f_g$ but are minimally influenced by $AoA$. 
$C_{Qc}$ plot shows a similar increasing trend but with an exception at $AoA$ of $2\si{\degree}$  (low $AoA$), which shows a little disparity as compared to that of the steady flow condition (the dotted line below). 
The gap between $C_{Qf}$ and $C_{Qc}$ can be attributed to the occurrence of ventilation hysteresis at each flow condition, which decreases with $f_g$ for $AoA\ 6\si{\degree}$ (moderate $AoA$) and $AoA\ 10\si{\degree}$ (high $AoA$ hereafter).

In an attempt to adduce a possible mechanism for the observed $C_{Qf}$ trends, further investigation is conducted with high-speed imaging of the foamy cavity state slightly below the $C_{Qf}$ for both steady and unsteady flow conditions (Fig.~\ref{fig4}a). 
As shown from a sample image, the light source from the other side of the test section illuminates the bubbly flow, and the bubble movements result in intensity fluctuations in the recorded images. 
By taking the standard deviation of the entire image sequence for a sufficiently long duration at each pixel location, it is possible to provide an estimate of the size of the recirculation region beyond the cavitator. 
The recirculation region consists of a high density of bubbles that are optically thick (and therefore shows a relatively low standard deviation of the pixel intensity), and the remaining flow field exhibits large fluctuations in intensity at the rear location. 
A timespan of four times the gust cycle is considered sufficient for the computation of standard deviation of intensity variations in the images.

Figure~\ref{fig4} illustrates the methodology adopted to trace out the recirculation region at the supercavity rear portion. 
The schematic shown in Fig.~\ref{fig4}a outlines the region of interest chosen for the analysis of the bubble images. 
Standard deviation values at each pixel location for steady (Fig.~\ref{fig4}b) flows, unsteady (Fig.~\ref{fig4}c) flow conditions, and the absolute difference between the two (Fig.~\ref{fig4}d) are presented. 
The gray dotted regions in Fig.~\ref{fig4}b and c (representing the inferred boundary of the recirculation region for steady incoming flow case) indicates that the recirculation region for the steady flow is more elongated than that for the unsteady flow. 
Stated differently, Fig.~\ref{fig4}d suggests that the incoming flow unsteadiness imposes fluctuation of the recirculation region beyond the cavitator. 
Therefore, a slight increment in $C_{Qf}$ for unsteady flows may be attributed to the vertical fluctuations of the recirculation region, which adversely affects the bubble coalescence efficiency, by decreasing the chance of bubble collisions to form a larger bubble in the low $Fr$ regime. 
Similarly, an increase in $AoA$ has shown to amplify the magnitude of the vertical perturbation \cite{shao2018POF} of the incoming flow, and therefore may adversely impact overall bubble coalescence efficiency. 
Likewise, for the same $AoA$, a greater $f_g$ means there is lesser time interval conceded per gust cycle for the bubbles to coalesce and form a supercavity, and hence $C_{Qf}$ grows with $f_g$.

\begin{figure*}%[pos=b]
	\centering
	\includegraphics[scale=1]{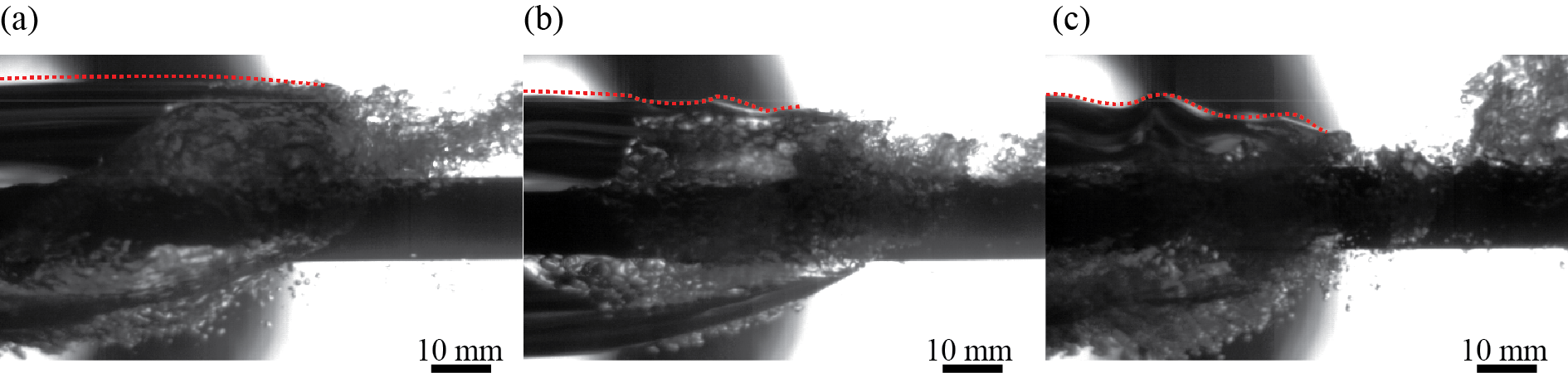}
	\caption{High-speed imaging taken near the closure for low $Fr$ regime with various $AoA$, (a) $2\si{\degree}$, (b) $6\si{\degree}$, and (c) $10\si{\degree}$, respectively. Red dotted lines help visualize the air-water interface distortion due to the development of instability. Stronger instability is observed for higher $AoA$ that leads to periodic additional gas leakage. For low $AoA$, however, no additional gas leakage is observed.}
	\label{fig5}
\end{figure*}

High-speed images acquired near the closure of the supercavity also provide interesting visual evidence on the $C_{Qc}$ trends as well. 
From Fig.~\ref{fig3}, it is observed that $C_{Qc}$ plot shows an increasing trend with $f_g$ for moderate and high $AoA$ but has a minimal significance for low $AoA$. 
Figure~\ref{fig5} clearly demonstrates that the interface instability at the upper surface near the closure is periodically developed at each gust cycle when the flapping foil is heading upward. 
Note that Shao et al. \cite{shao2018POF} attributed such phenomenon to the inverse alignment of the density gradient with respect to the direction of gravity. 
For low $AoA$, a wavy pattern of the overall shape of the supercavity is observed without any noticeable distortion of the air-water interface (Fig.~\ref{fig5}a). 
This instability, however, appears and becomes stronger for unsteady flows with higher $AoA$ (Fig.~\ref{fig5}b-c), periodically shedding off a certain volume of gas near the closure which eventually leads to a periodic fluctuation of cavity length due to the additional gas leakage. 
As regards the effect of $f_g$ on $C_{Qc}$ trends, similar to the formation case, an increase in $f_g$ essentially implies lesser time available per gust cycle for the supercavity to recover its length caused by the periodic shedding of the gas pocket (for moderate and high $AoA$). 
High-speed imaging of the supercavity collapse process further suggests the role of interface instability that has not been discussed before in the explanations of its trend on BFM cavitator provided by Karn et al. \cite{karn2016bETFS}, where the change of $C_{Qc}$ for unsteady flows was attributed only to the internal pressure fluctuation in a supercavity. 

\subsection{Ventilation demands in high $Fr$ regime}\label{sec:highfr}

\begin{figure}%[pos=b]
  \centering
  \includegraphics[scale=1]{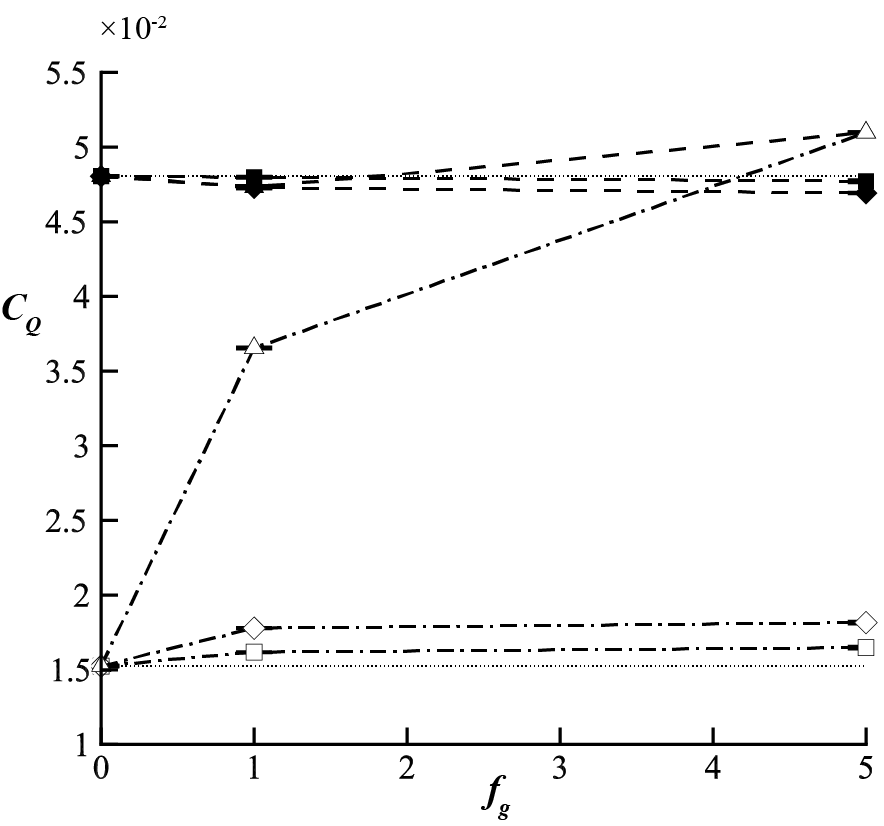}
  \caption{Ventilation coefficients at high $Fr$ regime under varying flow unsteadiness modified by $AoA$ and $f_g$. Dotted horizontal lines indicate steady flow condition, dashed lines indicate $C_{Qf}$ trends, dashed-dotted lines indicate $C_{Qc}$ trends. Symbols: $\blacksquare$, $C_{Qf}$ at $AoA\ 2\si{\degree}$; $\square$, $C_{Qc}$ at $AoA\ 2\si{\degree}$; $\blacklozenge$, $C_{Qf}$ at $AoA\ 6\si{\degree}$; $\lozenge$, $C_{Qc}$ at $AoA\ 6\si{\degree}$; $\blacktriangle$, $C_{Qf}$ at $AoA\ 10\si{\degree}$; $\triangle$, $C_{Qc}$ at $AoA\ 10\si{\degree}$.}
\label{fig6}
\end{figure}

In contrast to the observations in the low $Fr$ regime, the $C_Q$ plots for the high $Fr$ regime, as shown in Fig.~\ref{fig6}, exhibit a slightly different trend, both qualitatively and quantitatively. 
In particular, for low and moderate $AoA$, $C_{Qf}$ plot shows a marginal reduction with a consequent rise in $f_g$, and the slopes become steeper for the higher $AoA$. 
$C_{Qc}$ plot still shows a rising trend even for the lowest $AoA$. 
However, akin to the low $Fr$ case, the ventilation hysteresis gap reduces with increasing $AoA$ and $f_g$.

\begin{figure*}
  \centering
  \includegraphics[scale=1]{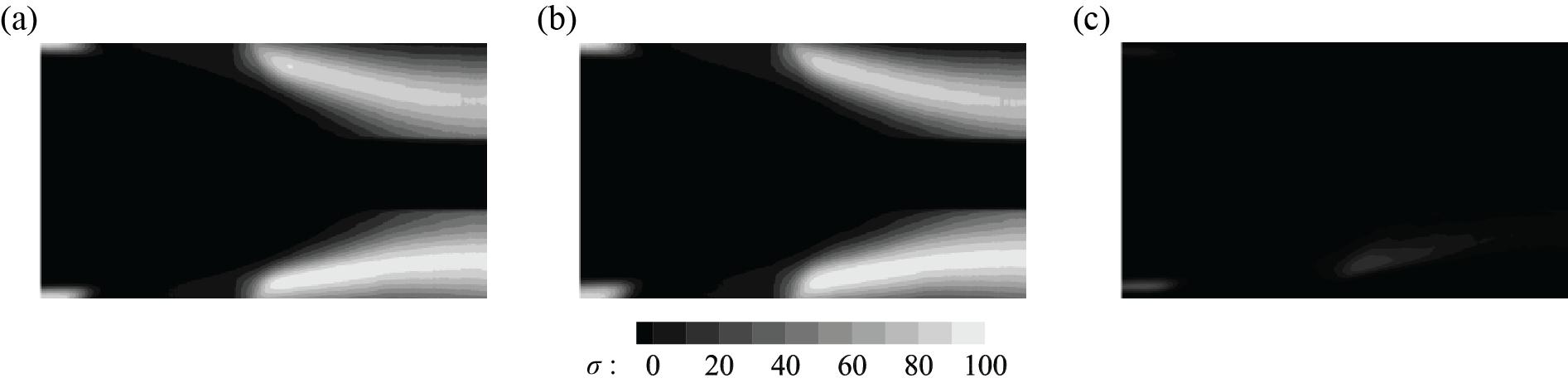}
  \caption{Standard deviations of image intensity for the (a) steady and (b) unsteady flows at high $Fr$ regime are plotted at each pixel location. Bright region indicates fluctuation of brightness throughout the recorded sequence. Difference between steady and unsteady flows are plotted in (c), which is not noticeable.}
\label{fig7}
\end{figure*}

Next, in a manner analogous to the low $Fr$ regime case, the recirculation regions are traced out at the supercavity rear portion by computing standard deviations of each pixel intensity in the recorded images of foamy for steady (Fig.~\ref{fig7}a) and unsteady (Fig.~\ref{fig7}b) flow conditions, to understand the $C_{Qf}$ trends observed in the high $Fr$ regime at low and moderate $AoA$. 
Surprisingly, as shown from Fig.~\ref{fig7}c, the recirculation region in the high $Fr$ regime remained relatively unchanged with the incoming flow unsteadiness. 
In this regime, the inherent high-level turbulence near the wake region may minimize the effect of vertical flow perturbation. 
Also, it has been reported that the flow unsteadiness results in higher transverse Reynolds stress, or $\nu'\nu'/U_M^2$, within the recirculation region beyond the backward-facing step or blunt body \cite{yoshioka2001IJHFF,konstantinidis2008IJHFF}. 
The amplified vertical velocity perturbations are imposed on the bubble motion, enhancing the bubble collision probability that is crucial to the coalescence process within the confined recirculation region, which is already constricted by the presence of the mounting strut. 
Thus, in the situation of the presence of severe flow unsteadiness, $C_{Qf}$ decreases in comparison to the steady flow case. 
Further, a greater $AoA$ entails imposing a consequently larger vertical perturbation to the incoming flow \cite{shao2018POF} and it is possible that the resulting vertically intensified  bubble movements may be responsible for the reduction of $C_{Qf}$ for higher $AoA$. 
Higher $f_g$ has a similar role in contributing to the increased vertical perturbation of the bubble movements for the given low and moderate $AoA$ unsteady flow cases. 

\begin{figure*}%[pos=b]
	\centering
	\includegraphics[scale=1]{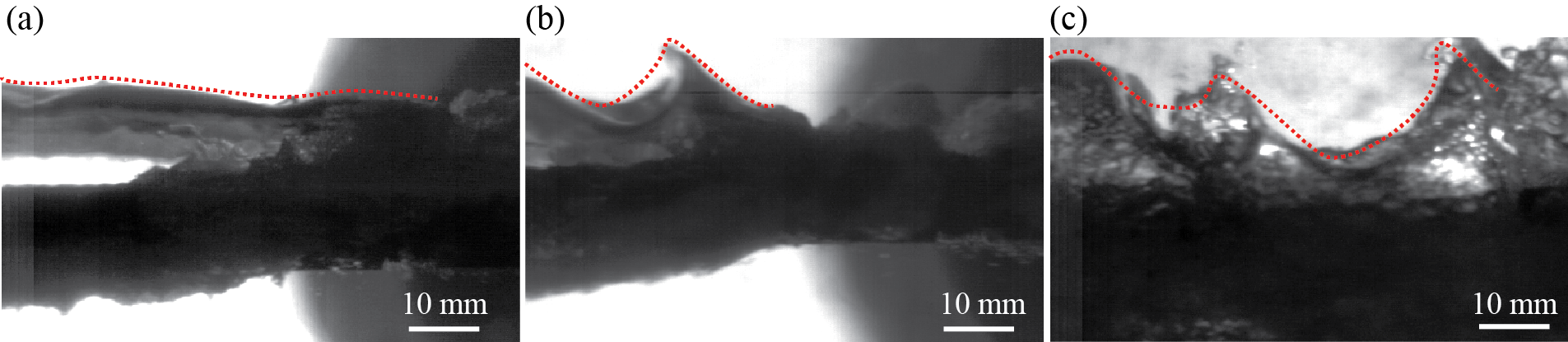}
	\caption{Samples of high-speed images acquired near the closure for high $Fr$ regime with various $AoA$, (a) $2\si{\degree}$, (b) $6\si{\degree}$, and (c) $10\si{\degree}$, respectively. Red dotted lines help visualize the air-water interface distortion due to the development of interface instability. Stronger instability is observed for higher $AoA$ that leads to periodic additional gas leakage.}
	\label{fig8}
\end{figure*}

The influence of flow unsteadiness on $C_{Qc}$ for this specific unsteady flow conditions (low and moderate $AoA$) are further plotted in Fig.~\ref{fig8}a-b. 
As shown from the figure, even for the lowest $AoA$, a certain gas volume sheds off from the supercavity due to the developed interface instability. 
In fact, compared to the low $Fr$ regime, all $AoA$ cases show stronger distortion of the air-water interface. 
As a result, $C_{Qc}$ required for the supercavity to sustain itself off collapsing, increases owing to the additional gas leakage. 
The effect of $AoA$ and $f_g$ on $C_{Qc}$ seems to share the same mechanism as that of the low $Fr$ regime. 

\begin{figure*}%[pos=b]
	\centering
	\includegraphics[scale=1]{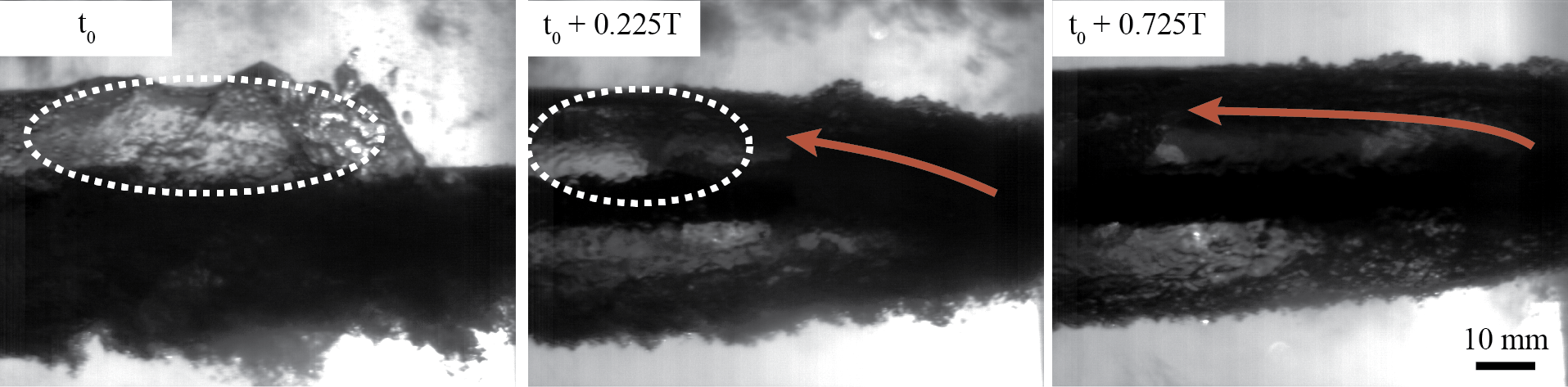}
	\caption{Snapshots of high-speed imaging taken near the closure for high $Fr$ at the highest $AoA$ ($10\si{\degree}$) and $f_g$ ($5\ \mathrm{Hz}$). As the cavity recovers its length from its shortest ($t=t_0$), the re-entrant jet becomes stronger. Such re-entrant jet blocks the light path inside the supercavity and the internal region correspondingly becomes darker, and the highlighted transparent regions decrease correspondingly. Red arrows indicate the re-entrant jet.}
	\label{fig9}
\end{figure*}

For the high $AoA$, the trends observed for both $C_{Qf}$ and $C_{Qc}$ exhibit a significantly disparate behavior as compared to the rest of the experimental conditions. 
For instance, as shown from Fig.~\ref{fig6}, $C_{Qf}$ slightly diminishes at low $f_g$ but then surges again to a value that is greater than that for the steady flow condition. 
In addition, $C_{Qc}$ drastically grows with $f_g$ and exceeds the steady $C_{Qf}$ and finally coincides with the $C_{Qf}$ trend at higher $f_g$. 
Such a phenomenon suggests the absence of ventilation hysteresis for highly unsteady flows with respect to $AoA$ and $f_g$. 
At this condition, as shown from the high-speed images (Fig.~\ref{fig9}), a drastic change of the cavity length is observed for every gust cycle. 
In particular, when the flapping hydrofoil reaches its amplitude during every gust cycle, the cavity length is truncated to a very short one, due to the strong interface instability (Fig.~\ref{fig8}c). 
This is followed by a recovery of cavity elongation in sync with the augmentation in re-entrant jet intensity, from its weakest to the strongest. 
As presented in Fig.~\ref{fig9}, the re-entrant jet becomes strong enough to block the light path from the illumination source located on the other side of the test section. 
As a result, the optical thickness of the internal cavity grows as the re-entrant jet strengthens, despite the presence of the supercavity with a smooth surface. 
The absence of the ventilation hysteresis at this flow condition may be caused by this phenomenon, as the majority of the internal cavity consists of water and there exists less volume of gas inside. 
Such phenomenon is consistent with the observation from Kawakami and Arndt \cite{kawakami2011JFE} where they noted that FFM configuration supercavity has less ventilation hysteresis gap compared to that of the BFM supercavity, due to the presence of the mounting strut that reduces the internal gas volume for the FFM models.

\subsection{Further insights into the collapse air entrainment}\label{sec:Shed}

\begin{figure}%[pos=b]
	\centering
	\includegraphics[scale=1]{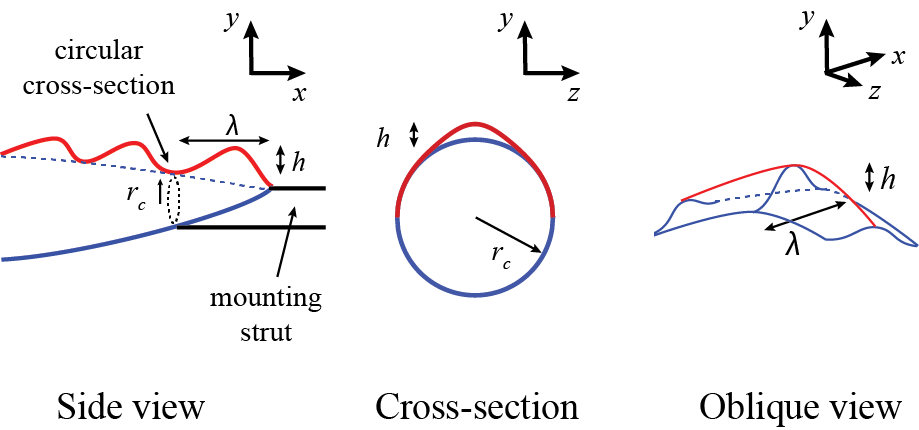}
	\caption{Illustration of interface instability characterization. Interface instability is characterized by its wavelength ($\lambda$), wave amplitude ($h$), and wave frequency. The interface instability near the closure of the supercavity is assumed to be 2D along the perimeter of the upper half of the cross-section.}
	\label{fig10}
\end{figure}

The alterations in $C_{Qc}$ due to the flow unsteadiness is not quantitatively understood yet. 
Such a difference may be attributed to various reasons such as internal pressure fluctuation, the re-entrant jet momentum change (similar to \cite{shao2020ETFS}), or the additional gas leakage induced by the interfacial instability. 
However, we observed that the flow unsteadiness and its state (e.g. flapping foils heading upward or downward) has a negligible effect on the re-entrant jet-speed estimation (following an approach similar to \cite{shao2020ETFS}), indicating that the re-entrant jet momentum does not govern the cavity collapse under unsteady flows. 
It is quite likely, then to conjecture, of the predominant influence that the interfacial instability may have on $C_{Qc}$. 
Hence, to substantiate this hypothesis, high-speed images of unsteady flow are investigated by estimating the additional gas leakage from the observed interface instability for both high and low $Fr$ regimes. 

Figure~\ref{fig10} illustrates how the interface instability is characterized for calculating the individual volume of gas bubbles shedding off from the supercavity. 
The interface instability is approximated as a sinusoidal pattern which is characterized by its wavelength, wave amplitude, and wave frequency. 
Based on the observation that the instability is stronger when $AoA$ is higher, we further approximate the cross-sectional shape of the instability as sinusoidal since the interface instability is associated with the normal interaction between the vertical component of the free stream velocity and the upper interface of the supercavity. 
Here, we assume that the cross-sectional shape of the supercavity is circular under unperturbed flow conditions. 
The volume of gas shedding off the supercavity is calculated based on such approximations. 
It is important to note that the additional shedding only occurs at every gust cycle when the flapping foil is heading downward and also that the wave frequency mentioned here is independent of the gust frequency. 
Thus, the additional gas leakage per second induced by the interface instability is calculated by multiplying the individual shedding volume with half of the wave frequency. 
The acquired leakage rate is in volumetric terms, so an additional step is required to convert it into a mass flowrate. 
Assuming the internal pressure is spatially uniform, cavitation number from \cite{shao2020ETFS} is used for mass flowrate estimation. 
We further assume that the internal pressure is temporally uniform, since the internal pressure fluctuation range has been reported to be less than $1\ \mathrm{kPa}$ \cite{shao2018POF} which yields a maximum uncertainty of less than $2\%$ when converting the volumetric flow rate into the mass flow rate.

\begin{figure}%[pos=b]
	\centering
	\includegraphics[scale=1]{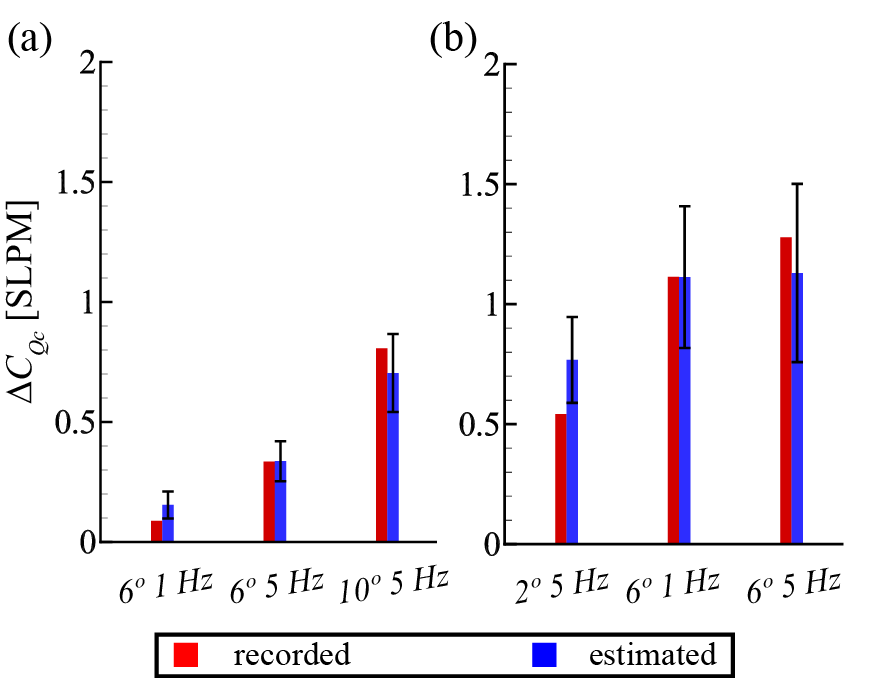}
	\caption{Increments in collapse ventilation demand due to incoming flow unsteadiness for (a) low and (b) high $Fr$ regimes. Red bars indicate the recorded data from Fig.~\ref{fig3} and Fig.~\ref{fig6} and blue bars indicate the calculated differences.}
	\label{fig11}
\end{figure}

An estimation of the gas leakage rates under different experimental conditions can now be obtained. 
Figure~\ref{fig11} presents a bar plot depicting the variations of $\Delta C_{Qc}$ due to flow unsteadiness ($C_{Qc,unsteady}-C_{Qc,steady}$), across a range of $f_g$ and $AoA$ both in the low and high $Fr$ regimes. 
It is worth noting that in this depiction, instances of extreme unsteadiness are avoided such as $AoA\ 2\si{\degree}$ cases at low $Fr$ when the interface instability does not lead to any additional gas leakage. 
Likewise, for $AoA\ 10\si{\degree}$ cases at high $Fr$, the supercavity truncates its length so much in each gust cycle that it precludes any reasonable estimation of wave characteristics. 
The red bars in the figure are the actual increment of $C_{Qc}$ in SLPM recorded in the current study, corresponding to the $C_{Qc}$ data points in Fig.~\ref{fig3} and Fig.~\ref{fig6}. 
The blue bars indicate the calculated additional gas leakage rate induced by interface instability. 
As shown from the figure, the calculated leakage rates show an overestimation for the low $AoA$ cases and an underestimation for high $AoA$ cases. 
Such a difference may be attributed to the sinusoidal assumption of the interface profile. 
In reality, the shape of the surface instability varies from cnoidal to breaking waves depending on the wave amplitudes. 
Also, interface instability characterization is based on the high-speed imaging near the closure, and uncertainty in the estimation of the wave amplitude of the weak instability is tantamount to greater uncertainties in the calculations of gas leakage rate. 
Nevertheless, considering that $10\ \mathrm{cm^3/s}$ at standard conditions is roughly equivalent to $0.6$ SLPM, our estimation shows a reasonable match with the experimentally measured values. 
Thus, our results indicate that the additional gas leakage induced by the interface instability indeed governs the increase in $C_{Qc}$ observed for unsteady flows.

\section{Summary and Conclusion}\label{sec:Conclusions}
In this study, we have investigated the ventilation characteristics of a supercavity generated by a forward-facing cone-type cavitator under unsteady flows. 
Flow unsteadiness is adjusted by changing either the angle of attack ($AoA$) or the frequency ($f_g$) of the flapping foils located upstream of the test section. 
At a lower free stream velocity, the formation ventilation demand ($C_{Qf}$) and collapse ventilation demand ($C_{Qc}$) both increase with flow unsteadiness except for the collapse demand at the lowest $AoA$. 
At a larger free stream velocity however, the observed trends are found to be markedly different. 
With increase in flow unsteadiness, a gradual rise in $C_{Qc}$ is observed up to an $AoA$ of $6\si{\degree}$. 
But, at a higher $AoA$ of $10\si{\degree}$, the collapse demand shows a drastic escalation and coincides with the formation demand trend, implying that the ventilation hysteresis no longer exists in highly unsteady flows. 
High-speed imaging reveals that the change in the recirculation region behind the cavitator with flow unsteadiness is responsible for the change in $C_{Qf}$. 
At a lower free stream velocity, the recirculation region after the cavitator shifts vertically in response to the incoming waves induced by the gust generator. 
At a higher free stream velocity, the recirculation region shows a negligible disparity between the steady and unsteady flow conditions. 
An increased transverse Reynolds stress may be ascribed to the decreased $C_{Qf}$ due to the higher collision and coalescence probability of the dispersed bubbles within the recirculation region. 
For $C_{Qc}$, it is believed that the increasing trend with incoming flow unsteadiness is due to the interface instability developed at the upper surface of the supercavity near the closure, periodically shedding off the bubbles from the supercavity. 
Subsequently, the additional gas leakage rate is estimated based on the high-speed imaging of the interface instability. 
The estimated additional gas leakage rate shows a reasonable match with the measurement, suggesting that the interface instability governs the increase in $C_{Qc}$ for unsteady flows. 

Our measurements provide a detailed explanation of the change of $C_Q$ with incoming flow unsteadiness, which sheds some light on the ventilation strategy of cavitating vehicles. 
Specifically, the ventilation demand trends under unsteady flows show some difference with Karn et al. \cite{karn2016bETFS} where they measured the ventilation demands with a backward-facing step cavitator, which bears significantly less semblance to supercavitating vehicles that have a solid body inside the supercavity. 
Our study suggests that depending upon the flow unsteadiness, the underlying physics that govern the formation and collapse of a supercavity are distinctly different. 
Further, by approximating the shape and profile of the interface instability as sinusoidal and assuming the cross-section of the supercavity to be circular, an estimate of the shed-off gas volume rate from the supercavity is calculated. 
These estimations however may yield a higher uncertainty when the wave amplitude increases. 
In reality, the equations of the curvilinear supercavity profile are more complex, particularly with the continuously varying amplitude. 
In addition, in the current study, we have only investigated one flow condition in both low and high $Fr$ regime. 
For supercavitating vehicles in practice, different ventilation strategies should be applied as the ventilation demand trends vary with vehicle speeds. 
Therefore, it would be of great interest to further examine the change of the ventilation demands with varying free stream velocity to investigate the $Fr$ effect, in order to comprehensively understand the underlying physics.

\section*{Declaration of Competing Interests}
The authors declare that they have no known competing financial interests or personal relationships that could have appeared to influence the work reported in this paper.

\section*{Acknowledgements}
This work is supported by the Office of Naval Research (Program Manager, Dr. Thomas Fu) under grant No. N 000141612755. 
%\appendix
%\section{My Appendix}
%Appendix sections are coded under \verb+\appendix+.
\printcredits
%\verb+\printcredits+ command is used after appendix sections to list 
%author credit taxonomy contribution roles tagged using \verb+\credit+ 
%in frontmatter.


\bibliographystyle{model1-num-names}
%\bibliographystyle{cas-model2-names}
%\bibliographystyle{unsrt}

\bibliography{cas-refs}

\begin{thebibliography}{22}
\expandafter\ifx\csname natexlab\endcsname\relax\def\natexlab#1{#1}\fi
\providecommand{\url}[1]{\texttt{#1}}
\providecommand{\href}[2]{#2}
\providecommand{\path}[1]{#1}
\providecommand{\DOIprefix}{doi:}
\providecommand{\ArXivprefix}{arXiv:}
\providecommand{\URLprefix}{URL: }
\providecommand{\Pubmedprefix}{pmid:}
\providecommand{\doi}[1]{\href{http://dx.doi.org/#1}{\path{#1}}}
\providecommand{\Pubmed}[1]{\href{pmid:#1}{\path{#1}}}
\providecommand{\bibinfo}[2]{#2}
\ifx\xfnm\relax \def\xfnm[#1]{\unskip,\space#1}\fi
%Type = Book
\bibitem[{Franc and Michel(2006)}]{franc2006book}
\bibinfo{author}{J.-P. Franc}, \bibinfo{author}{J.-M. Michel},
  \bibinfo{title}{Fundamentals of cavitation}, volume~\bibinfo{volume}{76},
  \bibinfo{publisher}{Springer science \& Business media},
  \bibinfo{year}{2006}.
%Type = Techreport
\bibitem[{May(1975)}]{may1975report}
\bibinfo{author}{A.~May}, \bibinfo{title}{Water entry and the cavity-running
  behavior of missiles}, \bibinfo{type}{Technical Report}, Navsea
  Hydroballistics Advisory Committee Silver Spring Md, \bibinfo{year}{1975}.
%Type = Article
\bibitem[{Sanabria et~al.(2014)Sanabria, Balas, and Arndt}]{escobar2015thesis}
\bibinfo{author}{D.~E. Sanabria}, \bibinfo{author}{G.~Balas},
  \bibinfo{author}{R.~Arndt},
\newblock \bibinfo{title}{Modeling, control, and experimental validation of a
  high-speed supercavitating vehicle},
\newblock \bibinfo{journal}{IEEE Journal of Oceanic Engineering}
  \bibinfo{volume}{40} (\bibinfo{year}{2014}) \bibinfo{pages}{362--373}.
%Type = Book
\bibitem[{Nesteruk(2012)}]{Nesteruk2012Book}
\bibinfo{author}{I.~Nesteruk}, \bibinfo{title}{Supercavitation: Advances and
  Perspectives A collection dedicated to the 70th jubilee of Yu. N. Savchenko},
  \bibinfo{publisher}{Springer Science \& Business Media},
  \bibinfo{year}{2012}.
%Type = Article
\bibitem[{Karn et~al.(2016{\natexlab{a}})Karn, Arndt, and Hong}]{karn2016aJFM}
\bibinfo{author}{A.~Karn}, \bibinfo{author}{R.~E. Arndt},
  \bibinfo{author}{J.~Hong},
\newblock \bibinfo{title}{An experimental investigation into supercavity
  closure mechanisms},
\newblock \bibinfo{journal}{Journal of Fluid Mechanics} \bibinfo{volume}{789}
  (\bibinfo{year}{2016}{\natexlab{a}}) \bibinfo{pages}{259--284}.
%Type = Article
\bibitem[{Karn et~al.(2016{\natexlab{b}})Karn, Arndt, and Hong}]{karn2016bETFS}
\bibinfo{author}{A.~Karn}, \bibinfo{author}{R.~E. Arndt},
  \bibinfo{author}{J.~Hong},
\newblock \bibinfo{title}{Gas entrainment behaviors in the formation and
  collapse of a ventilated supercavity},
\newblock \bibinfo{journal}{Experimental Thermal and Fluid Science}
  \bibinfo{volume}{79} (\bibinfo{year}{2016}{\natexlab{b}})
  \bibinfo{pages}{294--300}.
%Type = Techreport
\bibitem[{Semenenko(2001)}]{semenenko2001report}
\bibinfo{author}{V.~N. Semenenko}, \bibinfo{title}{Artificial supercavitation.
  physics and calculation}, \bibinfo{type}{Technical Report}, Ukranian Academy
  of Sciences, Kiev Inst of Hydromechanics, \bibinfo{year}{2001}.
%Type = Article
\bibitem[{Ahn et~al.(2010)Ahn, Lee, and Kim}]{ahn2010IJNAOE}
\bibinfo{author}{B.~K. Ahn}, \bibinfo{author}{C.~S. Lee},
  \bibinfo{author}{H.~T. Kim},
\newblock \bibinfo{title}{Experimental and numerical studies on
  super-cavitating flow of axisymmetric cavitators},
\newblock \bibinfo{journal}{International Journal of Naval Architecture and
  Ocean Engineering} \bibinfo{volume}{2} (\bibinfo{year}{2010})
  \bibinfo{pages}{39--44}.
%Type = Article
\bibitem[{Moghimi et~al.(2017)Moghimi, Nouri, and Molavi}]{moghimi2017JAFM}
\bibinfo{author}{M.~Moghimi}, \bibinfo{author}{N.~Nouri},
  \bibinfo{author}{E.~Molavi},
\newblock \bibinfo{title}{Experimental investigation on supercavitating flow
  over parabolic cavitators},
\newblock \bibinfo{journal}{Journal of Applied Fluid Mechanics}
  \bibinfo{volume}{10} (\bibinfo{year}{2017}) \bibinfo{pages}{95--102}.
%Type = Inbook
\bibitem[{Ahn et~al.(2018)Ahn, Jeong, and Park}]{ahn2018CAV}
\bibinfo{author}{B.~K. Ahn}, \bibinfo{author}{S.~W. Jeong},
  \bibinfo{author}{S.~T. Park}, \bibinfo{title}{An experimental investigation
  of artificial supercavitation with variation of the body shape},
  \bibinfo{year}{2018}. \DOIprefix\doi{10.1115/1.861851_ch12}.
%Type = Article
\bibitem[{Shao et~al.(2020)Shao, Balakrishna, Yoon, Li, Liu, and
  Hong}]{shao2020ETFS}
\bibinfo{author}{S.~Shao}, \bibinfo{author}{A.~Balakrishna},
  \bibinfo{author}{K.~Yoon}, \bibinfo{author}{J.~Li}, \bibinfo{author}{Y.~Liu},
  \bibinfo{author}{J.~Hong},
\newblock \bibinfo{title}{Effect of mounting strut and cavitator shape on the
  ventilation demand for ventilated supercavitation},
\newblock \bibinfo{journal}{Experimental Thermal and Fluid Science}
  \bibinfo{volume}{118} (\bibinfo{year}{2020}) \bibinfo{pages}{110173}.
%Type = Article
\bibitem[{Lee et~al.(2013)Lee, Kawakami, and Arndt}]{lee2013JFE}
\bibinfo{author}{S.~J. Lee}, \bibinfo{author}{E.~Kawakami},
  \bibinfo{author}{R.~E. Arndt},
\newblock \bibinfo{title}{Investigation of the behavior of ventilated
  supercavities in a periodic gust flow},
\newblock \bibinfo{journal}{Journal of Fluids Engineering}
  \bibinfo{volume}{135} (\bibinfo{year}{2013}).
%Type = Article
\bibitem[{Karn et~al.(2015)Karn, Arndt, and Hong}]{karn2015ETFS}
\bibinfo{author}{A.~Karn}, \bibinfo{author}{R.~E. Arndt},
  \bibinfo{author}{J.~Hong},
\newblock \bibinfo{title}{Dependence of supercavity closure upon flow
  unsteadiness},
\newblock \bibinfo{journal}{Experimental Thermal and Fluid Science}
  \bibinfo{volume}{68} (\bibinfo{year}{2015}) \bibinfo{pages}{493--498}.
%Type = Article
\bibitem[{Shao et~al.(2018)Shao, Wu, Haynes, Arndt, and Hong}]{shao2018POF}
\bibinfo{author}{S.~Shao}, \bibinfo{author}{Y.~Wu},
  \bibinfo{author}{J.~Haynes}, \bibinfo{author}{R.~E. Arndt},
  \bibinfo{author}{J.~Hong},
\newblock \bibinfo{title}{Investigation into the behaviors of ventilated
  supercavities in unsteady flow},
\newblock \bibinfo{journal}{Physics of Fluids} \bibinfo{volume}{30}
  (\bibinfo{year}{2018}) \bibinfo{pages}{052102}.
%Type = Article
\bibitem[{Jiang et~al.(2018)Jiang, Shao, and Hong}]{jiang2018POF}
\bibinfo{author}{Y.~Jiang}, \bibinfo{author}{S.~Shao},
  \bibinfo{author}{J.~Hong},
\newblock \bibinfo{title}{Experimental investigation of ventilated
  supercavitation with gas jet cavitator},
\newblock \bibinfo{journal}{Physics of Fluids} \bibinfo{volume}{30}
  (\bibinfo{year}{2018}) \bibinfo{pages}{012103}.
%Type = Article
\bibitem[{Wu et~al.(2019)Wu, Liu, Shao, and Hong}]{wu2019JFM}
\bibinfo{author}{Y.~Wu}, \bibinfo{author}{Y.~Liu}, \bibinfo{author}{S.~Shao},
  \bibinfo{author}{J.~Hong},
\newblock \bibinfo{title}{On the internal flow of a ventilated supercavity},
\newblock \bibinfo{journal}{Journal of Fluid Mechanics} \bibinfo{volume}{862}
  (\bibinfo{year}{2019}) \bibinfo{pages}{1135--1165}.
%Type = Article
\bibitem[{Yoon et~al.(2020)Yoon, Qin, Shao, and Hong}]{yoon2020EXIF}
\bibinfo{author}{K.~Yoon}, \bibinfo{author}{S.~Qin}, \bibinfo{author}{S.~Shao},
  \bibinfo{author}{J.~Hong},
\newblock \bibinfo{title}{Internal flows of ventilated partial cavitation},
\newblock \bibinfo{journal}{Experiments in Fluids} \bibinfo{volume}{61}
  (\bibinfo{year}{2020}) \bibinfo{pages}{1--15}.
%Type = Article
\bibitem[{Kopriva et~al.(2008)Kopriva, Arndt, and Amromin}]{kopriva2008JFE}
\bibinfo{author}{J.~Kopriva}, \bibinfo{author}{R.~E. Arndt},
  \bibinfo{author}{E.~L. Amromin},
\newblock \bibinfo{title}{Improvement of hydrofoil performance by partial
  ventilated cavitation in steady flow and periodic gusts},
\newblock \bibinfo{journal}{Journal of Fluids Engineering}
  \bibinfo{volume}{130} (\bibinfo{year}{2008}).
%Type = Article
\bibitem[{Karn et~al.(2016)Karn, Shao, Arndt, and Hong}]{karn2016cETFS}
\bibinfo{author}{A.~Karn}, \bibinfo{author}{S.~Shao}, \bibinfo{author}{R.~E.
  Arndt}, \bibinfo{author}{J.~Hong},
\newblock \bibinfo{title}{Bubble coalescence and breakup in turbulent bubbly
  wake of a ventilated hydrofoil},
\newblock \bibinfo{journal}{Experimental Thermal and Fluid Science}
  \bibinfo{volume}{70} (\bibinfo{year}{2016}) \bibinfo{pages}{397--407}.
%Type = Article
\bibitem[{Yoshioka et~al.(2001)Yoshioka, Obi, and Masuda}]{yoshioka2001IJHFF}
\bibinfo{author}{S.~Yoshioka}, \bibinfo{author}{S.~Obi},
  \bibinfo{author}{S.~Masuda},
\newblock \bibinfo{title}{Turbulence statistics of periodically perturbed
  separated flow over backward-facing step},
\newblock \bibinfo{journal}{International Journal of Heat and Fluid Flow}
  \bibinfo{volume}{22} (\bibinfo{year}{2001}) \bibinfo{pages}{393--401}.
%Type = Article
\bibitem[{Konstantinidis and Balabani(2008)}]{konstantinidis2008IJHFF}
\bibinfo{author}{E.~Konstantinidis}, \bibinfo{author}{S.~Balabani},
\newblock \bibinfo{title}{Flow structure in the locked-on wake of a circular
  cylinder in pulsating flow: Effect of forcing amplitude},
\newblock \bibinfo{journal}{International Journal of Heat and Fluid Flow}
  \bibinfo{volume}{29} (\bibinfo{year}{2008}) \bibinfo{pages}{1567--1576}.
%Type = Article
\bibitem[{Kawakami and Arndt(2011)}]{kawakami2011JFE}
\bibinfo{author}{E.~Kawakami}, \bibinfo{author}{R.~E. Arndt},
\newblock \bibinfo{title}{Investigation of the behavior of ventilated
  supercavities},
\newblock \bibinfo{journal}{Journal of Fluids Engineering}
  \bibinfo{volume}{133} (\bibinfo{year}{2011}).

\end{thebibliography}


\end{document}